\documentclass[prl, twocolumn,notitlepage,superscriptaddress,citeautoscript,showpacs,amsmath,amstex,amssymb,mathfonts,showpacks]{revtex4-1}

\pdfoutput=1

\usepackage{amsmath}
\usepackage{amssymb}
\usepackage{graphicx}
\usepackage{appendix}
\usepackage{enumerate}
\usepackage{setspace}
\usepackage{mathrsfs}
\usepackage{color}
\usepackage[protrusion=true,expansion=true]{microtype}

\usepackage{hyperref}
\hypersetup{
	pdftoolbar=true,        		 
	pdfmenubar=true,     		 
	pdffitwindow=false,     	 
	pdfstartview={FitH},    	 
	pdftitle={Logarithmically slow relaxation in quasi-periodically driven random spin chains},   
	pdfauthor={Dumitrescu, Vasseur, Potter},     		 
	pdfsubject={},   			 
	pdfcreator={},   	        	   	 
	pdfproducer={}, 		
	pdfnewwindow=true,      	
	colorlinks=true,       		
	linkcolor=black,          	
	citecolor=blue,        		
	filecolor=magenta,    		
	urlcolor=blue          		
}

\newcommand{\bx}{\mathbf{x}}

\newcommand{\bhx}{\hat{\textbf{x}}}
\newcommand{\bhy}{\hat{\textbf{y}}}

\newcommand{\ud}{\mathrm{d}}

\newcommand{\vect}[1]{\textbf{#1}}

\newcommand{\OA}{\Omega_{0}}
\newcommand{\OB}{\Omega_{1}}

\newcommand{\<}{\langle}

\renewcommand{\>}{\rangle}
\renewcommand{\(}{\left(}
\renewcommand{\)}{\right)}
\renewcommand{\[}{\left[}
\renewcommand{\]}{\right]}

\newcommand{\CC}[2]{{\left[ #1, #2 \right]}}

\newcommand{\header}[1]{\noindent{\bf#1 -- }}

\usepackage{ulem}

\begin{document}

\author{Philipp T. Dumitrescu}
\email{philippd@utexas.edu}
\affiliation{Department of Physics, University of Texas at Austin, Austin, TX 78712, USA}

\author{Romain Vasseur}
\affiliation{Department of Physics, University of California, Berkeley, CA 94720, USA}
\affiliation{Materials Science Division, Lawrence Berkeley National Laboratories, Berkeley, CA 94720, USA}
\affiliation{Department of Physics, University of Massachusetts, Amherst, MA 01003, USA}

\author{Andrew C. Potter}
\affiliation{Department of Physics, University of Texas at Austin, Austin, TX 78712, USA}

\title{Logarithmically Slow Relaxation in Quasi-Periodically Driven Random Spin Chains}

\date{\today}

\begin{abstract}
We simulate the dynamics of a disordered interacting spin chain subject to a quasi-periodic time-dependent drive, corresponding to a stroboscopic Fibonacci sequence of two distinct Hamiltonians. Exploiting the recursive drive structure, we can efficiently simulate exponentially long times. After an initial transient, the system exhibits a long-lived glassy regime characterized by a logarithmically slow growth of entanglement and decay of correlations analogous to the dynamics at the many-body delocalization transition. Ultimately, at long time-scales, which diverge exponentially for weak or rapid drives, the system thermalizes to infinite temperature. The slow relaxation enables metastable dynamical phases, exemplified by a ``time quasi-crystal'' in which spins exhibit persistent oscillations with a distinct quasi-periodic pattern from that of the drive. We show that in contrast with Floquet systems, a high-frequency expansion strictly breaks down above fourth order, and fails to produce an effective static Hamiltonian that would capture the pre-thermal glassy relaxation.
\end{abstract} 

\keywords{}
\pacs{}

\maketitle

\normalem

\header{Introduction}
Interacting quantum many-body systems often exhibit chaotic dynamics that rapidly scramble quantum information and lead to highly entangled states whose local properties are thermal and classical~\cite{PhysRevA.43.2046, PhysRevE.50.888}. A dramatic exception occurs in isolated and disordered systems where many-body localization (MBL)
can arrest thermalization, resulting in quantum coherent dynamics at arbitrarily high energy density~\cite{2014arXiv1404.0686N,Altman:2015aa,1742-5468-2016-6-064010}.
This dichotomy naturally raises fundamental questions about when and how a system thermalizes. What are the universal features governing the dynamical approach to the final -- thermal or non-thermal -- state? More practically, what classes of protocols allow one to manipulate a many-body system without rapidly scrambling its stored quantum information?

Given their large bandwidth and dense spectrum, one might naively expect that \emph{any} persistent dynamical manipulation of an isolated, interacting quantum many-body system leads to runaway heating to a featureless infinite-temperature state. Indeed, random time-dependent manipulations have recently been shown to cause rapid growth of entanglement, accompanied by universal hydrodynamic features~\cite{2016arXiv160806950N,2017arXiv170508910V,2017arXiv170508975N}. 
However, this expectation is violated in time-periodically driven (Floquet) systems with strong disorder, in which sufficiently rapid driving maintains MBL and indefinitely avoids heating~\cite{PhysRevLett.115.030402,PhysRevLett.114.140401,ABANIN20161}. Even in the absence of disorder, rapid periodic driving leads to long-lived pre-thermal phenomena~\cite{PhysRevB.95.014112,PhysRevLett.115.256803,KUWAHARA201696,PhysRevLett.116.120401,PhysRevX.7.011026, weidinger2017floquet-prether,chandran2016interaction-sta,canovi2016stroboscopic-pr,bukov2015prethermal-floq, gopalakrishnan2017noise-induced-subdiffusion}. 
 Floquet-MBL systems have been shown to exhibit remarkable dynamic phenomena from spontaneous time-translation symmetry breaking~\cite{PhysRevLett.116.250401,PhysRevB.94.085112,PhysRevLett.117.090402,PhysRevLett.118.030401,Zhang:2017aa,Choi:2017aa} to dynamical topological phases with no equilibrium analog~\cite{PhysRevB.82.235114, PhysRevLett.106.220402, PhysRevX.3.031005, PhysRevLett.116.250401,PhysRevB.93.245145,PhysRevB.93.201103,PhysRevX.6.041001,PhysRevB.94.125105,PhysRevX.6.041070, PhysRevLett.118.115301, 2017arXiv170101440P}. 

The stark contrast between the behaviors under random and periodic driving can be understood by a simple argument: local time-dependent Hamiltonians can only make local rearrangements. In strongly disordered systems, such rearrangements have a non-zero energy cost and are generically non-resonant with harmonics of the driving frequency. This heuristic forms the basis for more sophisticated considerations for the stability of Floquet-MBL systems~\cite{ABANIN20161}, which are supported by numerical simulations~\cite{PhysRevLett.115.030402,PhysRevLett.114.140401}, and cold-atom experiments~\cite{Bordia:2017aa}. Using similar arguments, one can rule out the stability of MBL to random time-dependent drives, which have continuous frequency spectra capable of resonantly inducing arbitrary local transitions leading to thermalization. 

\begin{figure*}
\centering
\includegraphics[width=\textwidth]{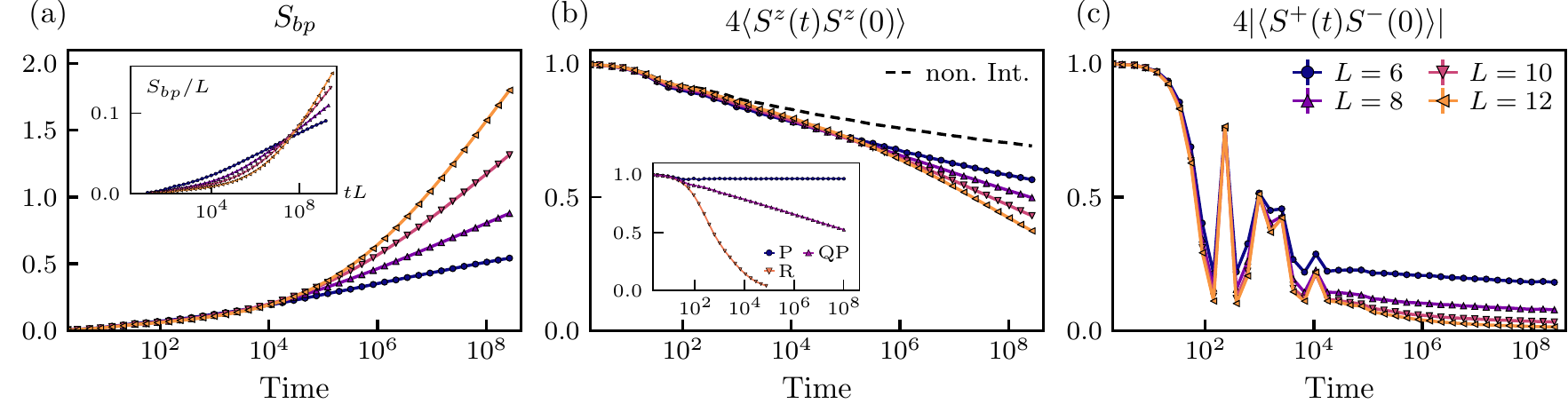}
\caption{{{\bf Quasi-Periodically Driven Spin Chain. -- } \label{fig:MainPlot} 
Time evolution under the quasi-periodic driving sequence, with $J_0 = 0, \delta J = \pi / 30, \lambda = 1$ and varying $L$ (markers defined in (c)). All quantities are averaged over states in the global $S^z = 0$ sector of the spin chain and averaged over at least 3000 disorder realizations. (a)~The bipartite entanglement $S_{bp}(t)$. Inset:~The normalized entanglement  $S_{bp} / L$. (b)~Onsite correlation function $C^{zz}(t)$ on site $i = L / 2$. This plot additionally shows (dashed line) the case of driving an $L=150$ chain in the non-interacting limit of \eqref{eq:ham}; see Supplemental Material \cite{SupMat}. Inset:~Comparison of driving with periodic (P), quasi-periodic (QP) and random (R) sequences of the elementary unitaries, with $L=8$. The random case is averaged over $20$ different random sequences, each with $100$ disorder realizations. (c)~Correlation function $C^{+-}(t)$ on site $i = L / 2$. 
}}
\vspace{-0.2in}
\end{figure*}

In this paper, we consider an intermediate case between periodic and random driving by subjecting a strongly disordered quantum many-body system to a drive with quasi-periodic time-dependence. The quasi-periodic drive has a dense, but sharply discontinuous frequency spectrum that occupies a set of measure zero. A priori, it is not clear whether the density of spectral content will drive heating and thermalization or whether its sparsity will preserve MBL. We find that quasi-periodic driving does eventually lead to thermalization to a featureless infinite temperature state, but only after a long time $t_\text{th}$ that grows exponentially in the inverse driving strength and the rate of driving. While reminiscent of pre-thermalization in delocalized Floquet systems~\cite{PhysRevB.95.014112,PhysRevLett.115.256803,KUWAHARA201696,PhysRevLett.116.120401,PhysRevX.7.011026}, the dynamics before $t_\text{th}$ are not described by an effective finite temperature equilibrium.
Instead, this regime shows a logarithmically slow relaxation of correlations and growth of entanglement, which we will call glassy dynamics. This glassy behavior is analogous to the critical dynamics at the transition between MBL and thermal systems in non-driven settings~\cite{Vosk:2015aa,Potter:2015aa,PhysRevX.5.041047}. 
We explore to what extent the quasi-periodic evolution can be reduced to an effective static Hamiltonian, connecting our study to the question of reducibility of differential equations with quasi-periodic coefficients \cite{jauslin1991spectral-quasiperiodic, blekher1992floquet-spectrum}. The glassy relaxation regime can host new metastable dynamical phases, which we illustrate with a quasi-periodic analog of time-translation symmetry breaking -- a ``time quasi-crystal".

\header{Model}
To address the fate of a quantum many-body system under quasi-periodic driving, we numerically simulate spin-1/2 chains, subjected to a stroboscopic drive consisting of a Fibonacci sequence of unitary evolutions:
\begin{align}
U_n &= U_{n-2}U_{n-1},
\label{eq:fibo}
\end{align}
for $n \geq 2$. The sequence is initialized by two elementary unitaries formed from two different static Hamiltonian evolutions: $U_0 = \exp\({-i\lambda H_+}\)$ and $U_1 = \exp\({-i\lambda H_-}\)$, where  
\begin{align}
H_{\pm} = \sum_{i=1}^{L} h_i S^z_i +  \sum_{i=1}^{L-1}\(J_0\pm \delta J\)\vect{S}_i\cdot\vect{S}_{i+1}.\label{eq:ham}
\end{align}
The $h_i$ are random fields drawn independently for each site from a uniform distribution $h \in [-2\pi, 2\pi)$, $J_0$ is a static interaction, $\delta J$ represents the strength of the quasi-periodic driving and $\lambda \in [0,1]$ is the characteristic driving time-scale.   We will focus on the regime $\left\vert J_0\pm \delta J \right\vert \lesssim 1.7$, where $H_\pm$  as static Hamiltonians would be MBL~\cite{Luitz}. As such, they are separately described by emergent local integrals of motion (LIOM) with definite $S^z$ value \cite{PhysRevLett.111.127201}. Unless otherwise noted, we will take $J_0 = 0$.
An appealing feature of the recursive nature of the drive is that it enables simulation of exponentially long Fibonacci times $t_n = F_{n+1} \sim \varphi^{n+1}$ with only $n$ unitary multiplications; here $\varphi = (1+\sqrt{5})/2$ is the golden ratio. This enables us to simulate the long-time physics, limited only by machine precision.

\header{Results}
We focus on three observables: the $z$-component of spin $C^{zz}(t) = 4 \<S_i^z(t)S_i^z(0)\>$, whose total value is conserved by the evolution, and whose local dynamics are related to spin-transport, the transverse spin-fluctuations $C^{+-}(t) = 4 \vert \< S_i^+(t)S_i^-(0)\>\vert$, which encodes the dephasing of quantum superpositions of up and down spins, and the bipartite (half-system) entanglement entropy $S_{bp}(t)$.

Before discussing the results, we summarize the behavior of these quantities in static MBL, periodically driven (Floquet) MBL, and thermalizing systems. In a static or Floquet-MBL system, $C^{zz}(t)$ tends to a non-zero constant at long times, indicating the absence of spin-transport and  emergent conservation laws that produce infinite memory of the initial spin configuration~\cite{PhysRevLett.111.127201,PhysRevB.90.174202,PhysRevLett.115.030402,PhysRevLett.114.140401}. The transverse fluctuations $C^{+-}(t)$ decay as a power law in time from dephasing due to classical interactions among the local conserved quantities~\cite{PhysRevB.90.174302}. This dephasing also produces a logarithmically slow growth of entanglement $S_{bp}(t)\sim \log t$~\cite{PhysRevB.77.064426,PhysRevLett.109.017202,PhysRevLett.110.260601}.  On the other hand, in strongly thermal or randomly driven systems, the non-zero spin conductivity and chaotic scrambling leads to an exponential decay of correlation functions $C^{zz},C^{+-}\sim e^{-t/t_\text{th}}$ and a linear growth in $S_{bp}(t)\sim t$~\cite{1742-5468-2005-04-P04010,PhysRevLett.111.127205}. 
Finally, a clean delocalized system subject to rapid periodic driving exhibits a pre-thermalization regime, in which the system initially equilibrates with respect to an effective Hamiltonian at finite temperature. Pre-thermalization persist up to a time exponentially long in the driving frequency~\cite{PhysRevB.95.014112,PhysRevLett.115.256803,KUWAHARA201696,PhysRevLett.116.120401}, after which the system heats to a featureless infinite temperature state.

Figure \ref{fig:MainPlot} shows $C^{zz}$, $C^{+-}$, and $S_{bp}$ for quasi-periodic driving, in a quench from an initial product state. These observables are averaged over initial states and disorder realizations. We observe three distinct regimes: First, there is a short-time transient regime in which there is no distinction between periodic, quasi-periodic and random driving~(Fig.~\ref{fig:MainPlot}b inset). Next, there is a long-lived glassy relaxation regime where $S_{bp}$ grows and $C^{zz}$ decays logarithmically slowly. Finally, after a time-scale $t_\text{th}$ that is exponentially long for weak or rapid driving, the system ultimately heats up to infinite temperature with a non-zero rate, signaled by linear growth of entanglement and rapid decay of correlations. {Ultimately, $S_{bp}$ will saturate to its thermal value and $C^{zz}, C^{+-}$ decay to zero.}

The behavior of this quasi-periodic system is markedly distinct from the other scenarios mentioned above, as contrasted in the inset of Fig.~\ref{fig:MainPlot}b. Similar to an MBL system, $C^{+-}$ shows aperiodic oscillations that decay slowly. Unlike an MBL system, however, $C^{zz}$ does not saturate to a non-zero value. Taken together, these imply that the glassy relaxation regime does not possess LIOM. Nonetheless, it does not exhibit the rapid decay characteristic of a thermal system.

\begin{figure}[t]
\centering
\includegraphics[]{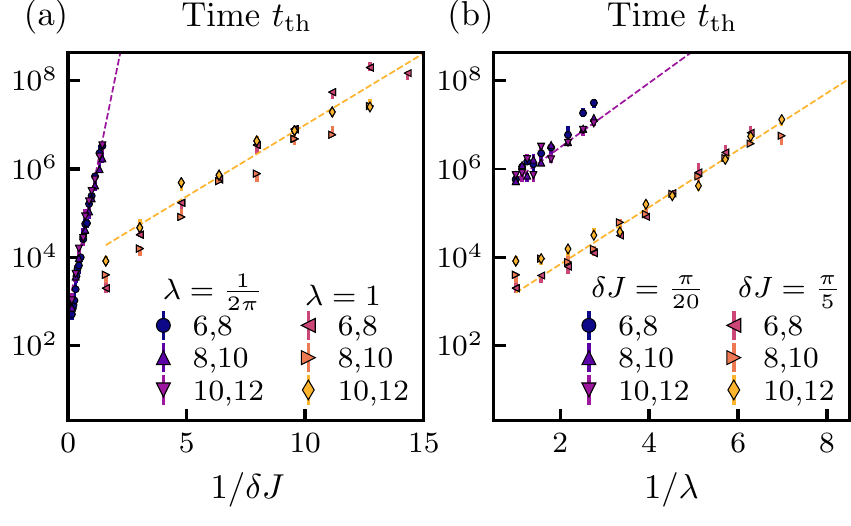}
\vspace{-2em}
\caption{{\bf Thermalization time $t_{\mathrm{th}}$. -- } \label{fig:parameterDependance} 
Thermalization time extracted from the crossing of $S_{bp} / L$, between pairs of $L$ $(6,8),(8,10), (10,12)$. (a) As a function of $1/\delta J$ for $\lambda =  1/2\pi, 1$ and (b) as a function of $\lambda$ for $\delta J =  \pi / 20, \pi / 5$. Error-bars are linear estimates in Fibonacci time; dashed lines are fits of form $\log t_n \sim 1/\lambda,  1/\delta J$ to the $(10,12)$ crossing.
}
\vspace{-0.2in}
\end{figure}

There are two ways we can identify the thermalization time $t_\mathrm{th}$: as the time where $C^{zz}$ curves of different $L$ separate from each other after the logarithmic decay or as the time where the normalized entanglement curves $S_{bp} / L$ cross at a single point as a function of $Lt$ 
(Fig.~\ref{fig:MainPlot}a inset). These two ways of extracting $t_\mathrm{th}$ follow each other closely and allow us to extract the parametric dependence of $t_\text{th}$ on $\delta J$ and $\lambda$ (Fig.~\ref{fig:parameterDependance})~\footnote{Since $t_\mathrm{th}$ is a cross-over scale, some observables show deviations earlier than others.}. At small $\lambda$ and $\delta J$ we find an asymptotic dependance which is consistent with
$
t_\text{th} \sim e^{1/\lambda}, t_\text{th} \sim e^{1/\delta J}, 
$
implying anomalously slow dephasing and decay over extremely long time-scales. At larger $\lambda, \delta J$ there may be deviations from this form.  In this respect, the logarithmic decay is reminiscent of the long-lived pre-thermal regime of non-MBL Floquet systems~\cite{PhysRevB.95.014112,PhysRevLett.115.256803,KUWAHARA201696,PhysRevLett.116.120401,PhysRevX.7.011026}. However, the entanglement growth in this region is slower than linear and consistent with logarithmic growth, which would not be the case of a system equilibrating to an  effective finite temperature and pre-thermal Hamiltonian. We note that such logarithmic decay is observed at the phase transition between MBL and thermal phases~\cite{Vosk:2015aa,Potter:2015aa,PhysRevX.5.041047}; here, we see this critical-like behavior without fine-tuning.

It is interesting to compare these results to those of a non-interacting analog of \eqref{eq:ham} (dashed line in Fig.~\ref{fig:MainPlot}b, {for detailed comparison} see \cite{SupMat}). The non-interacting system also exhibits a slow decay regime, but in this case there is no cross-over to fast thermalization ($t_\text{th}=+\infty$). This suggests that, despite the absence of local conserved quantities, the long lived glassy relaxation regime in the interacting case is nonetheless governed by {the dynamics of} emergent single-particle-like degrees of freedom.

\header{(Ir)reducibility of the quasi-periodic drive}
High-frequency expansions provide a useful tool for understanding pre-thermalization behavior in Floquet systems. They enable the computation of an effective static pre-thermal Hamiltonian and the expansion breakdown at long times indicates the onset of thermalization. Here, we attempt to develop a {generic} expansion of the {many-body} time-evolution operator organized in powers of $\lambda$ -- effectively a Magnus expansion -- taking advantage of the special self-similar structure of the Fibonacci drive.  Technical details are given in the Supplemental Material~\cite{SupMat}.

\begin{figure}
\centering
\includegraphics[]{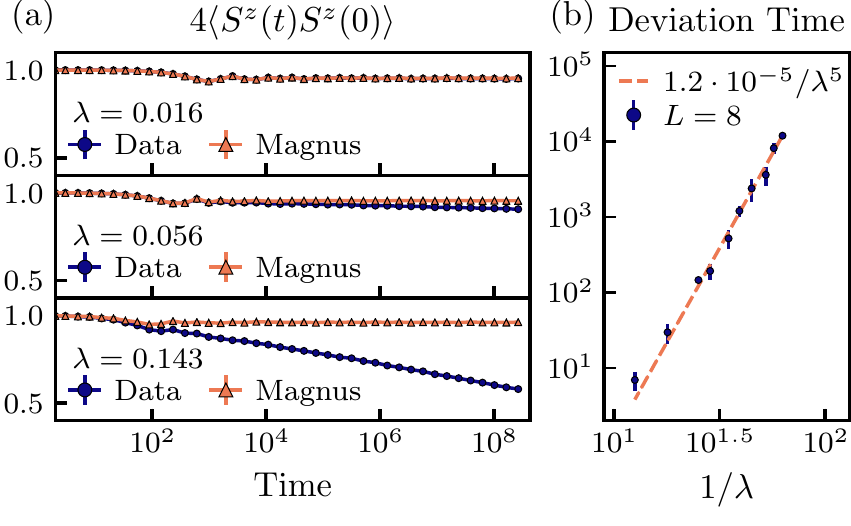}
\vspace{-2em}
\caption{{\bf Magnus Expansion. -- } \label{fig:magnus} 
(a) Onsite correlation function $C^{zz}(t)$ for  $L = 8, J_0 = 0, \delta J = \pi / 5$ and different $\lambda$ compared to that obtained by the Magnus expansion Hamiltonian at third order.
(b) Initial time at which $C^{zz}(t)$ of the Magnus expansion deviates by more than  $10^{-4}$ from the data.
}
\vspace{-0.2in}
\end{figure}

We can analytically construct a recursive Magnus expansion for $\Omega_{n} = \log U_n$, using the local deflation rule structure of quasiperiodic sequences \cite{Steinhardt1:1986,Steinhardt2:1986}. We can generate $U_{n+1}$ from $U_n$ by replacing $U_0 \to U_1$ and $U_1 \to U_0 U_1$ in the product defining $U_n$. We expand $\Omega_{n}$ onto a basis of nested commutators and construct and solve difference equations for the coefficients in this expansion, order-by-order in the degree $k$ of the commutator basis. Up to degree two: 
\begin{align*}
\Omega_{n} = F_{n-1} \OA + F_{n} \OB  + \tfrac{1}{2} \left\{ (-1)^{n} + F_{n-2}\right\}  \left[\OA,\OB\right].
\end{align*}
Explicit expressions for degrees $k=3, 4$ are given in the Supplemental Material~\cite{SupMat}.  In order to assign an effective static Hamiltonian interpretation, the asymptotic form for all coefficients need to be $\sim \varphi^{n}$, as above. However, for $k \geq 4$, the asymptotic behavior is $\sim \varphi^{(k-2)n}$. Therefore, the time where the non-Hamiltonian evolution dominates becomes increasingly short $t_n \sim \lambda^{-(k-1) / (k-3)}$. We note that this breakdown is fundamentally different from the breakdown of thermalization in the Floquet-Magnus case for periodic driving, which is due to a lack of convergence of the expansion.

Despite this, we find that truncating the expansion at $k = 3$ gives a Hamiltonian evolution which reproduces the data at small $\lambda$ remarkably well, with the exception of rare anomalous disorder configurations. Indeed, the time where this expansion deviates from the data scales with $\lambda^{-5}$, much later than the expected $\lambda^{-3}$  (Fig.~\ref{fig:magnus}). In no case, however, does the Magnus expansion capture the anomalous logarithmic decay of $C^{zz}$ or growth of $S_{bp}$ for $t<t_\text{th}$, suggesting these are inherently dynamical phenomena not governed by a static Hamiltonian, i.e.~not governed by an effective conserved (quasi)-energy. 

\header{Fibonacci time quasi-crystal}
The existence of an exponentially long lived quasi-MBL regime, with only logarithmically slow decay, raises the prospect of transient phases unique to quasi-periodically driven systems. These are analogous to metastable phases in pre-thermal Floquet settings, but with the important distinction that the quasi-periodically driven system does not require cooling to observe quantum coherent behavior.
To illustrate this possibility, we now construct a model that exhibits the quasi-periodic analog of discrete time-translation  symmetry-breaking~\cite{PhysRevLett.116.250401,PhysRevB.94.085112,PhysRevLett.117.090402,PhysRevLett.118.030401,Zhang:2017aa,Choi:2017aa}  -- a ``time quasi-crystal'' (TQC). The model uses the Fibonacci sequence of \eqref{eq:fibo}, but with elementary unitaries
\begin{align}
U_0 = e^{-i\theta \sum_i S_i^x},  U_1 = e^{-i\lambda \sum_i \(J_iS^z_iS^z_{i+1}+h^{z}_i S^z_i+h^{x}_i S^{x}_i\)}.
\label{eq:qtcmodel}
\end{align}
This model is closely inspired by the periodic version  introduced in~\cite{PhysRevLett.116.250401,PhysRevLett.117.090402}.

Consider the ideal case of \eqref{eq:qtcmodel}, where $\theta = \pi, h^{x}_{i} = 0$ and random $J_i, h^{z}_i$. Then $U_0 \sim \prod_i S_i^x$ applies a perfect, global spin-flip, while $U_1$ is made of only $S^z$ operators. A simple $S^z$-product state would merely acquire a phase under $U_1$ and flip under $U_0$. The time-evolution of a specific spin $\<S_i^z(t)S^z_i(0)\>$ exhibits an oscillating quasi-periodic pattern that is sharply distinct from the driving pattern. 
An elegant way to capture this difference is to view the quasi-periodic sequence as a projection of a 1d strip cutting through a regular 2d square lattice at an irrational angle (see \cite{SupMat}). The TQC spin response corresponds to a projection from a 2d lattice having a {doubled unit cell} compared to that for the drive.

\begin{figure}[t]
\centering
\includegraphics[]{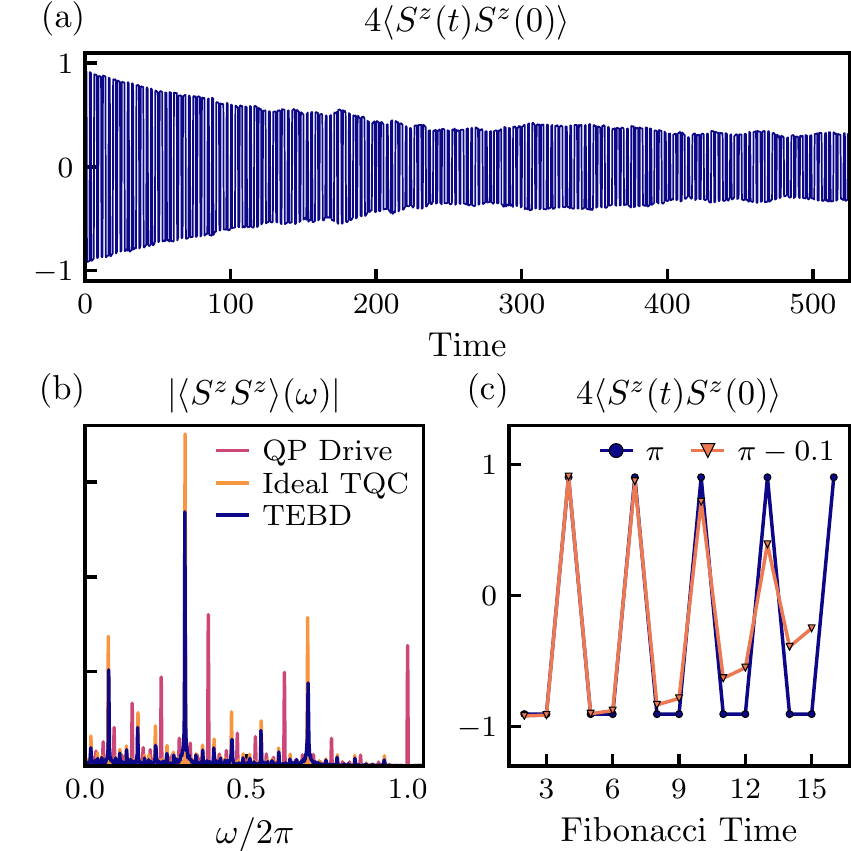}
\caption{{\bf Time Quasi-Crystal. -- } \label{fig:quasicrystal} 
(a) TEBD data of a single spin in a spin-1/2 chain subjected to drive \eqref{eq:qtcmodel} with $U_0$ occurring instantaneously. Parameters are $L = 60$, $\lambda = 1$, and $\theta = \pi - 0.1$ as well as random variables drawn from uniform distributions $J \in [2, 8] $, $h^{z}_{i} \in [0, 2] $, $h^{x}_{i} \in [0, 0.6]$. We show a single disorder realization.
(b) Fourier spectra of the quasi-periodic (QP) drive pattern, of the ideal TQC pattern and of the TEBD data.
(c) Magnetization at Fibonacci times, for an ideal ($\theta = \pi$) and a non-ideal ($\theta = \pi - 0.1$) pulse, shows period-3 oscillations characteristic of the TQC. 
}
\vspace{-0.2in}
\end{figure}

Alternatively, we can directly compare the Fourier spectrum of the spin response to that of the drive \cite{Steinhardt1:1986,Steinhardt2:1986}. For this, it is convenient to interpret $U_0$ in \eqref{eq:fibo} as arising from an instantaneous pulse, so that we can write the evolution in terms of a Hamiltonian with quasi-periodic delta-function ``kicking'':
$
H(t) = H_1 + \sum_{m=1}^M \delta\(t - t_m\) H_0,
$
where $t_m = \left\lfloor \varphi m \right\rfloor$ and $M$ is the largest integer such that $t_{M} \leq t$. 
In the ideal limit $\theta = \pi, h^{x}_{i} = 0$, the correlation function would satisfy  $\ud C^{zz} (t) / \ud t = 2 \sum_{m=1}^M  (-1)^m \delta\(t - t_m\)$. The spectrum of the spin-response is shifted compared to the drive (see Fig.~\ref{fig:quasicrystal} and Supplemental Material~\cite{SupMat}).  The distinction between the spin-response and drive patterns is even simpler if we consider stroboscopically measuring $C^{zz}(t)$ at Fibonacci times $t_n = F_n$. At these times, the initial spins have been flipped $F_{n-1}~\text{mod}~2$ times from their initial state. Since $F_k \mod 2$ form a repeating pattern with period $3$; the TQC is characterized by persistent period-$3$ oscillations in Fibonacci time.

These aspects also generalize straightforwardly to other time quasi-crystal phases. For example, we may replace the Ising spins ($\mathbb{Z}_2$)  by $N$-state clock spins ($\mathbb{Z}_N$)  in $U_1$ and replace $S^x$ by the operator that increments the clock spins in $U_0$ of \eqref{eq:qtcmodel}. In Fibonacci time, the spins would oscillate with the Pisano period $\pi(N)$; for $N=2,3,4,5$, $\pi(N)=3,8,6,20$. While the emergence of quasi-periodic correlations that have a different pattern from the drive can occur in ideally driven single spins \cite{feudel1995correlation-qp-spin}, this is special to fine-tuned drivings. In the many-body set-up \eqref{eq:qtcmodel}, the interactions give phase rigidity even away from the ideal limit $\theta = \pi$, as for a Floquet time-crystal \cite{PhysRevLett.118.030401}.

For $\theta\neq\pi$ or $h_x \neq 0$, the model becomes non-integrable and we lose analytic control. Figure~\ref{fig:quasicrystal} shows $C^{zz}(t)$ from time-evolving block decimation (TEBD)~\cite{PhysRevLett.91.147902,PhysRevLett.93.040502,SCHOLLWOCK201196} for system size $L=60$ starting from a product state. The TEBD calculations were done with Trotter step $0.01 \lambda$, keeping the discarded weight below $10^{-7}$ throughout the time evolution. Away from the ideal limit, the results largely track the ideal oscillations, but we clearly see the overall logarithmic decay in the quasi-periodic oscillations due to the quasi-MBL nature as discussed in the previous sections. In the Heisenberg chain \eqref{eq:ham} discussed above, the glassy relaxation was smoothly connected to the non-interacting limit. It is intriguing that this behavior is again observed in a system that is unconnected to any free fermion limit due to the longitudinal fields. This again suggests a possible description in terms of an emergent set of effectively single-particle, though non-conserved, degrees of freedom.

Despite that the system eventually thermalizes, for moderately small $\lambda$ the decay is sufficiently slow to permit many period-3 oscillations in Fibonacci time. This is a fundamentally different type of approximate non-equilibrium order than previously discussed for the cases of pre-thermal order in Floquet systems~\cite{PhysRevB.95.014112,PhysRevLett.115.256803,KUWAHARA201696,PhysRevLett.116.120401,PhysRevX.7.011026}, which require cooling to an effective prethermal ground state. 

Beyond this quasi-periodic generalization of a Floquet time-crystal, the slow  relaxation in the long-lived regime of glassy relaxation opens the door to more exotic quantum dynamical behavior such as long lived quasi-periodic topological phenomena. Investigating this intriguing possibility, and developing a systematic theoretical framework to characterize such metastable quantum phases will be an important challenge for future work.

\begin{acknowledgments}
\vspace{4pt}\noindent{\it Acknowledgements -- }
We thank M.~Kolodrubetz and N.Y.~Yao for insightful discussions. We especially thank S.~Gopalakrishnan for pointing out that the glassy relaxation also occurs in the non-interacting limit. Numerical simulations were performed at the Texas Advanced Computing Center (TACC) at the University of Texas at Austin. This work was supported by NSF DMR-1653007 (ACP) and the LDRD program at LBNL (RV). This work was performed in part at Aspen Center for Physics, which is supported by National Science Foundation Grant No.~PHY-1607611 (PTD \& ACP).
\end{acknowledgments}

\nocite{reutenauer1993free-lie-algebr}

\bibliography{qpdc-mbl-biblio}

\appendix
\onecolumngrid
\newpage
\section*{Supplementary material}
\twocolumngrid

\section{Non-Interacting Limit\label{sec:AppNonInt}}
\begin{figure}[b]
\centering
\includegraphics[]{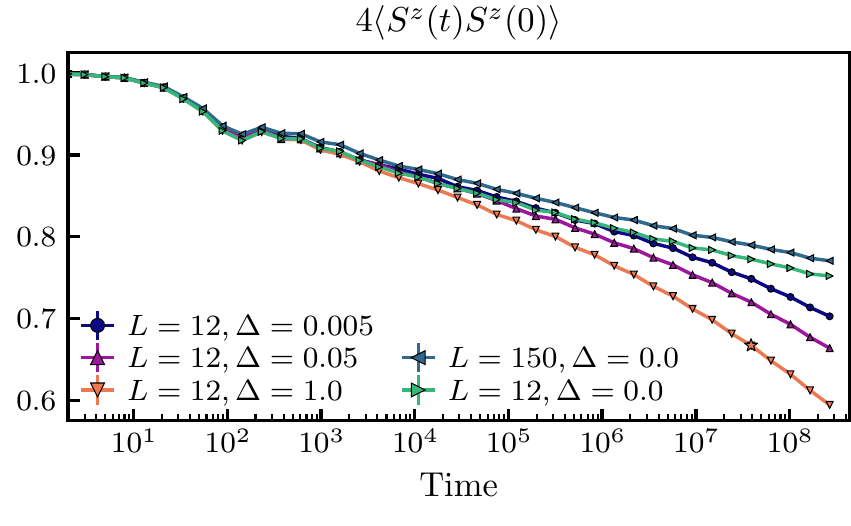}
\caption{{\bf Quasi-periodic evolution for varying interaction $\Delta$. -- } \label{fig:varyDeltaPlot} 
On-site correlation function, $C^{zz}(t)$, for  $J_0 = 0, \delta J = \pi / 5, \lambda =  2/ (5 \pi)$ and varying $\Delta$. The non-interacting data ($\Delta = 0$) is obtained by mapping the model to free fermions and has $5 \cdot 10^4$ disorder realizations; the curves are converged for system sizes $L \gtrsim 50$. 
Interacting models have $L=12$ and at least $4 \cdot 10^3$ disorder realizations. The star for $\Delta = 1.0$ indicates $t_\mathrm{th}$ as extracted from the entanglement entropy crossing; this time is off-scale for $\Delta = 0.005, 0.05$. 
}
\end{figure}

To adjust the strength of interactions, we generalize the Heisenberg term in \eqref{eq:ham} to an XXZ form:
\begin{align*}
H_{\pm} = \sum_{i=1}^{L} h_i S^z_i +  \sum_{i=1}^{L-1}J_\pm \left[ {S}^x_i{S}^x_{i+1} +  {S}^y_i{S}^y_{i+1} + \Delta  {S}^z_i{S}^z_{i+1} \right]
\end{align*}
with $J_\pm = J_0 \pm \delta J$. This model maps to fermions hopping in a disordered potential with interaction strength $\Delta$.

Figure~\ref{fig:varyDeltaPlot} shows the correlation function  $C^{zz}(t)$ for varying $\Delta$, which all show a long-lived glassy relaxation regime with  slow decay, though the rate of the logarithmic decay is renormalized by the interaction strength. For weak $\Delta$, the $C^{zz}(t)$ initially tracks the non-interacting curve, before the interactions come into effect and increase the rate of the logarithmic decay.

We see that the non-interacting dynamics also show slow relaxation, consistent with logarithmic or slower decay. For $\Delta = 1$ interaction effects appear essentially at the start of the logarithmic regime, while for $\Delta = 0.005, 0.05$ it occurs much later. 
The non-interacting systems do not rapidly thermalize, but rather continue their slow decay until $C^{zz}(t)$ approaches zero.
 
The observation that the glassy decay can be connected to the non-interacting limit shows an effective picture of quasi-periodically driven non-interacting degrees of freedom is still appropriate in this regime, despite the absence of bona fide local integrals of motion (LIOMs). Intriguingly, this behavior also appears in the time quasi-crystal models which do not map to a local fermionic model due to the longitudinal field, suggesting that this regime is described by a quasi-MBL like picture of driven LIOMs.

\section{Breakdown of the Magnus expansion}\label{sec:AppMagnus}
As discussed in the main text, we can analytically construct a recursive Magnus expansion for $\Omega_{n} = \log U_n$ order-by-order, using the local deflation rule of quasiperiodic sequences \cite{Steinhardt1:1986,Steinhardt2:1986}. If we write   $U_n$ as the quasi-periodic string of the elementary operators $U_{0},U_{1}$ and make the substitution $U_0 \to U_1$, $U_1 \to U_0 U_1$ we  generate the string for $U_{n+1}$; explicitly $U_{n+1}(U_{0}, U_{1}) = U_{n}(U_{1}, U_0 U_1)$. In terms of $\Omega$, this rule becomes 
\begin{align}
\Omega_{n+1}(\OA, \OB) &= \Omega_{n}(\OB, \OA * \OB), \label{eq:deflationOmegaPrime}
\end{align} 
where $\OA * \OB = \log(\exp\OA \exp{\OB})$, which we replace by the Baker-Campbell-Hausdorff (BCH) formula in our expansion, neglecting issues of the BCH convergence. We expand $\Omega_{n}$ onto a basis of commutators:
\begin{align*}
\Omega_{n}  &=   a_{n} \OA +  b_{n} \OB + h_{n} \Omega_{01}  + f_{n }\Omega_{001} + g_{n} \Omega_{011}  \\
&+ k_{1,n} \Omega_{0001} + k_{2,n} \Omega_{0011} + k_{3,n} \Omega_{0111} + \ldots
\end{align*}

\begin{figure}[b]
\centering
\includegraphics[]{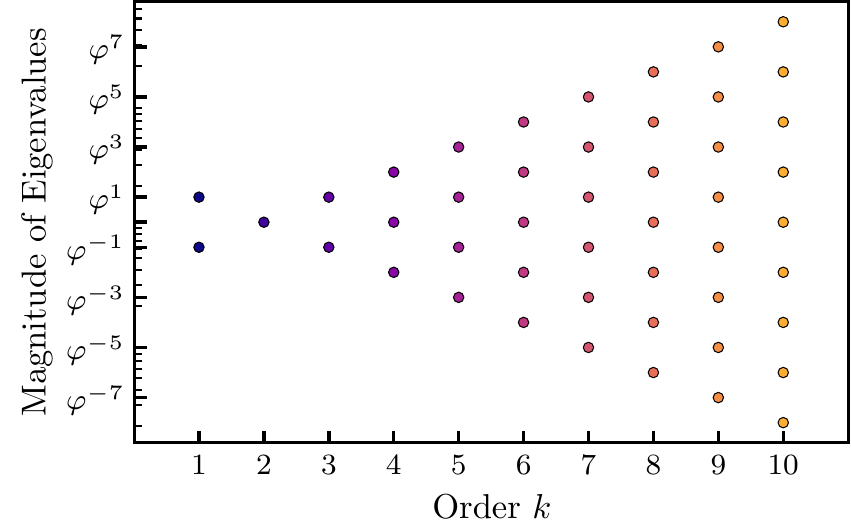}
\caption{{\bf Magnitude of Eigenvalues of $M$. -- } \label{fig:eigM} Absolute magnitude of the eigenvalues of $M$ in the recursion relation \eqref{eq:magnusCoeff} are given in terms of a simple pattern: at order $k$ they given by $\varphi^{m}$, where $m = - \vert k - 2 \vert, - \vert k  - 4 \vert, \ldots ,  \vert k - 2 \vert$. We see that for $k \geq 4$, the largest eigenvalues are $\geq \varphi$ and therefore the Hamiltonian interpretation breaks down.
}
\vspace{-0.15in}
\end{figure}

It is convenient to use the Lyndon basis \cite{reutenauer1993free-lie-algebr}:
\begin{align*}
&\OA, 										&&  \Omega_{011} = \CC{\CC{\OA}{\OB}}{\OB},\\
&\OB, 										&&  \Omega_{0001} =\CC{\OA}{\CC{\OA}{\CC{\OA}{\OB}}},	\\
&\Omega_{01} = \CC{\OA}{\OB}, 					&& \Omega_{0011} = \CC{\OA}{\CC{\CC{\OA}{\OB} }{\OB}},  		\\
&\Omega_{001} =\CC{\OA}{\CC{\OA}{\OB}}, 			&& \Omega_{0111} = \CC{\CC{\CC{\OA}{\OB}}{\OB}}{\OB}.	 
\end{align*}
Using \eqref{eq:deflationOmegaPrime} and the BCH formula, we find coupled difference equations for coefficients $\vect{a}$ of the commutators
\begin{equation}\label{eq:magnusCoeff}
\vect{a}_{n +1} = M \vect{a}_{n} + \vect{r}_{n}.
\end{equation}
Here $M$ is a matrix of integers, which solely arises from the substitution of $\OA \to \OB, \OB \to \OA + \OB$ in the commutators at the given degree $k$. The vector $\vect{r}_{n}$ contains coefficients only from lower degree terms and numerical factors from the BCH formula. This structure allows us to solve for the coefficients consecutively order-by-order. 

We are interested in the asymptotic behavior of $\vect{a}_{n}$ with $n$, since this determines whether we can assign a Hamiltonian interpretation to the expansion. Since \eqref{eq:magnusCoeff} is an inhomogenous linear difference equation, the asymptotic behavior is determined either by the largest eigenvalue of $M$ or by the asymptotic behavior of $\vect{r}_{n}$. The $k=1$ terms $\Omega_{n} = F_{n-1} \OA + F_{n} \OB$ define Fibonacci time
\begin{equation}
F_{n} = ({\varphi^{n} - \varphi^{-n}})/{\sqrt{5}}.
\end{equation}
An effective Hamiltonian would take the form $U_n\simeq e^{-iF_n H_\text{eff}}$, which can therefore be achieved only if he absolute value of the largest eigenvalues of all $M$ are $\leq \varphi$. If any eigenvalue of $M$ are larger than $\varphi$ and we write $U_n \simeq e^{-iF_n H_\text{eff}}$, the terms in $H_\text{eff}$ for a fixed $n$ would themselves grow without bound as $n$ is increased. In fact, we find a simple pattern of absolute value of the eigenvalues (Fig.~\ref{fig:eigM}). Not only does the Hamiltonian interpretation therefore breaks down at $k=4$, but higher orders become more important earlier in time.
\begin{widetext}
Finally, we give the coefficients up to $k=4$; note in particular the $\varphi^{2n}$ asymptotic form for $k=4$
\begin{align*}
a_{n} &= F_{n-1} , \qquad
b_{n} = F_{n} \\
h_{n} & = \frac{1}{2}\left[ (-1)^{n} + F_{n-2}\right],  \\
f_{n} &=  g_{n-1} + \frac{F_{n-1}}{12},	 \qquad
g_{n} = \frac{1}{12}\left\{F_{n-3} + (-1)^{n} \left[{2 F_{n} + F_{n-1}}{} - {3}{}\right] \right\},	\\
k_{1, n} &= \frac{(-1)^n}{120} \left\{ \varphi^{2n} (2 - \varphi) + \varphi^n \left[(-1)^n (4 \varphi - 7) - 2 - \varphi\right] + 2 \[5 + (-1)^n\] 
 + \frac{  (-1)^n(\varphi -3) - 4\varphi -3  }{\varphi^{n}}+   \frac{1 + \varphi }{\varphi^{2n}}  \right\}, \\
k_{2, n} &= \frac{(-1)^n}{120} \left\{ 2 \varphi^{2n} (1 - \varphi) + \varphi^n \left[(-1)^n (3 -  \varphi ) + 3 +4 \varphi\right] -  \[15 + 2 (-1)^n\] 
 + \frac{  (-1)^n(7 - 4\varphi) + \varphi + 2 }{\varphi^{n}}+   \frac{2 \varphi }{\varphi^{2n}}  \right\}, \\
k_{3, n} &= \frac{(-1)^n}{120} \left\{ \varphi^{2n}  +   \frac{1}{\varphi^{2n}} + \left(\varphi^n  + \frac{1}{\varphi^{n}} \right) \left[(-1)^n (3\varphi - 4 ) -1 -3 \varphi\right] + 2 \[5-   (-1)^n\] 
  \right\}.
\end{align*}
\end{widetext}
\newpage

\section{Fourier response of a Fibonacci time-crystal}

Here we briefly summarize the construction of the quasicrystal and TQC in terms of the projection method and given the explicit form of the Fourier spectrum of Figure \ref{fig:quasicrystal}. Following \cite{Steinhardt1:1986,Steinhardt2:1986}, we define a quasi-periodic sequence:
\begin{equation}\label{eq:qp1dseq}
x_{m} = m + \alpha + \frac{1}{\rho} \left\lfloor{\frac{m}{\sigma} + \beta} \right\rfloor,
\end{equation}
where $m$ is an integer, $\sigma$ is an irrational number and $\rho > 0$. Sequences with differing $\alpha, \beta$ are related to each other by a local isophormism and form a class. For the Fibonacci quasi-periodic sequence $\rho = \sigma = \varphi$ and the sequence consists two intervals $L, S$; in the truncated sequences of the unitary evolution \eqref{eq:fibo}, these intervals correspond to the unitaries $U_1, U_0$.

\begin{figure}[t]
\centering
\includegraphics[]{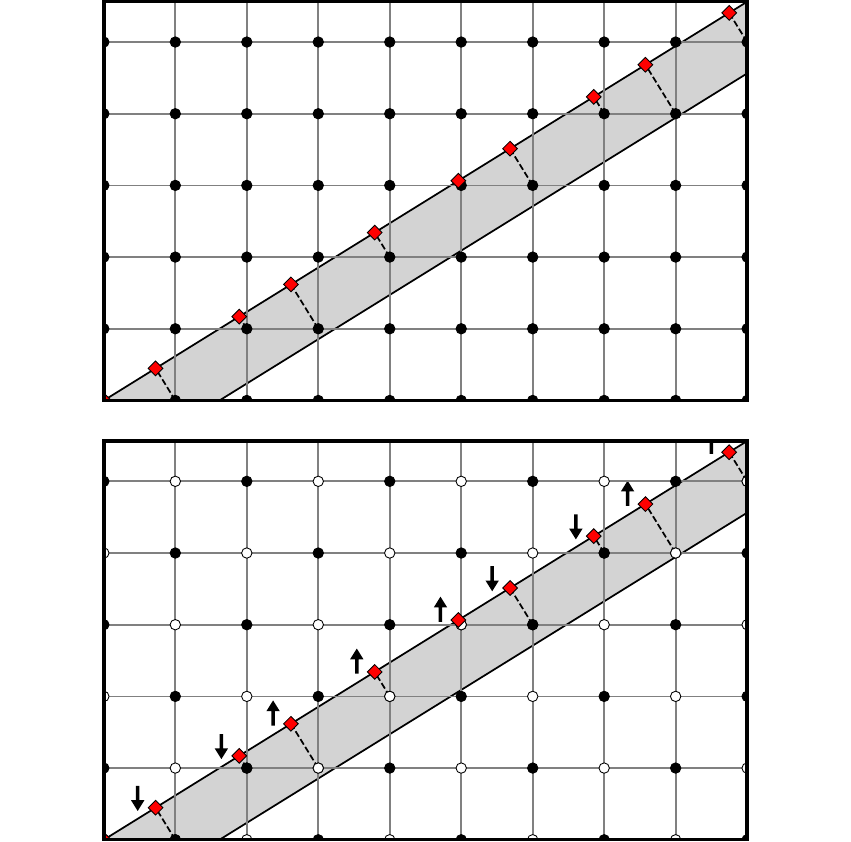}
\caption{{\bf Quasiperiodic and Time Quasi-Crystal Sequences from Projection. -- } \label{fig:fibproj} 
(upper) Projection method for the Fibonacci sequence $\rho = \sigma = \varphi$. The black points in the strip of length $1$ (shaded) on the 2d grid are projected onto the line $y_1 = x / \varphi$ to form the quasiperiodic sequence (red diamonds). The long $L$ and short $S$ intervals between points give the sequences of unitary evolutions \eqref{eq:fibo}.
(lower) For the time quasi-crystal, the Fibonacci pattern has an additional sign structure, shown as spins. During a $S$ interval the spin is flipped, while during a $L$ interval it is not; we can obtain this pattern by doubling the period of the underlying lattice in a checkerboard pattern as shown by the black and white dots.
}
\vspace{-0.2in}
\end{figure}

Many quasi-periodic sequences can be easily obtain by a projection from a two dimensional rectangular grid $\bx = a \bhx + b \bhy$ (Fig.~\ref{fig:fibproj}). To obtain the quasi-periodic sequences \eqref{eq:qp1dseq}, take all the lattice points between the two parallel lines
$
y_{1} = (\tan\theta) x + y_{0},  y_{2} = (\tan\theta) x + y_{0} - b,
$
and project them onto one of the lines, say $y_{1}$. The position along the line gives the desired sequence.  The identification of parameters is
$
a \cos \theta =1,  			 
 \beta = y_{0} / b,
b \sin \theta ={1}/{\rho} ,
\tan^{2} \theta = {1}/{(\sigma\rho)}
$.
The projection fixes $\alpha = \beta / \rho$; a change of $\alpha$ corresponds to a translation along $y_1$.

For the Fibonacci case shown in Fig.~\ref{fig:fibproj}, an $S$ intervals corresponds to a projections of a horizontal interval $\bhx$ onto $y_1$, while the $L$ interval corresponds to a projection of a diagonal interval  $\bhx + \bhy$. This allows us to formulate the time quasi-crystal (TQC) pattern as translation symmetry breaking of the original lattice. In the ideal TQC model \eqref{eq:qtcmodel}, the $L$ interval corresponds to the phase evolution $U_1$, whereas the $S$ interval corresponds to the spin flipping unitary $U_0$. We can therefore obtain the TQC pattern by signing vertices in a checkerboard pattern, doubling the unit cell of the lattice. We note that there are several ways to perform the projection, related to the symmetries of the sequence; in all cases, the associated TQC pattern has a doubled unit cell, although possibly with a different lattice symmetry breaking.

The Fourier transform of the delta-function train $f(x) = \sum_m \delta(x-x_m)$ of a quasi-periodic sequence, with $\alpha = \beta = 0$ for convenience, is  \cite{Steinhardt1:1986,Steinhardt2:1986}
\begin{equation*}
 f(k) =  \sum_{pq}\delta(k-k_{pq}) \frac{\sin\(X / 2\)}{X / 2} e^{i X / 2},
\end{equation*}
where $p,q$ are integers and
\begin{align*}
k_{pq} &= \frac{2 \pi \sigma \rho}{1 + \sigma \rho} \(p + \frac{q}{\sigma}\), & X &= \frac{2 \pi \sigma \rho}{1 + \sigma \rho} \(q - \frac{p}{\rho}\).
\end{align*}

In contrast, the TQC has the different quasiperiodic sequence
$
\widehat{f}(x) = \sum_{m} \[\delta(x - x_{2m}) - \delta(x - x_{2m +1}) \],
$
which has Fourier transform 
\begin{align*}
\widehat{f}(k)  =   \sum_{{p}q}\delta\(k-{\widehat{k}_{pq}}{}\) \frac{\sin\(\widehat{X} / 2\)}{\widehat{X} / 2} e^{i \widehat{X}/ 2},
\end{align*}
where
\begin{align*}
{\widehat{k}_{{p}q} }{} &= \frac{2 \pi \sigma \rho}{1 + \sigma \rho} \({p}-\frac{1}{2} + \frac{q}{\sigma}\), \\
\widehat{X} &= \frac{2 \pi \sigma \rho}{1 + \sigma \rho} \(q - \frac{p - 1/2}{\rho} \).
\end{align*}
We see that the new position of Fourier peaks and dominant amplitudes occur are a shift  for each the integer $p \to p - 1/2$ from the original quasiperiodic sequence. This shift from integers to half-integers exactly introduces lower frequency components. This is expected from the doubling of the unit cell in the projection construction and is another way to make precise the intuitive sense in which TQC sequence is ``longer'' than the quasi-periodic drive.

For the TEBD simulation, we applied the spin-flip unitary $U_0$ instantaneously. The time of application of the unitary $t_m$ forms a quasiperiodic sequence \eqref{eq:qp1dseq} with $\sigma = \varphi, \rho = 1, \alpha = 0, \beta =0$. The derivative of the correlation function $\ud C^{zz}(t) / \ud t$ follows the associated TQC quasi-periodic sequence. In Fig.~\ref{fig:quasicrystal}, we show the Fourier spectra of the quasi-periodic (QP) and time quasi-crystal (TQC) pattern, with an additional $1 / \omega$ amplitude factor to account for the derivative.

\begin{figure}[t]
\centering
\includegraphics[]{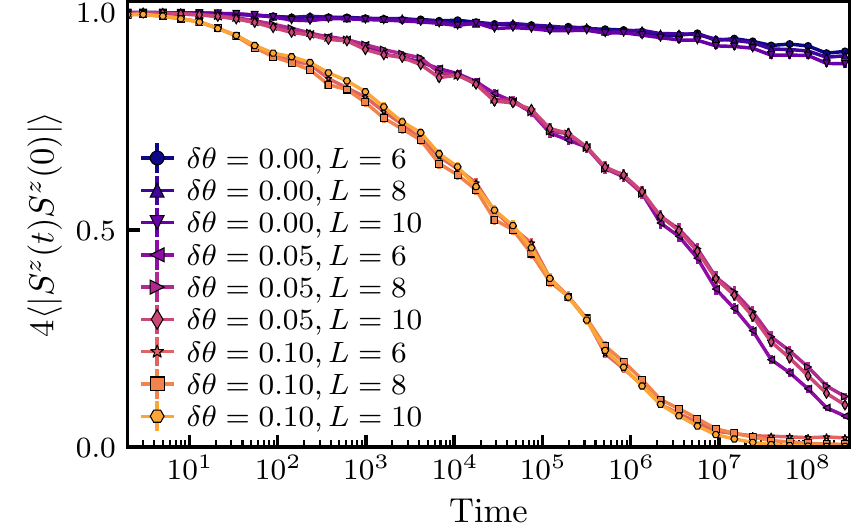}
\caption{{\bf Behavior of Time Quasi-Crystal at Long Times. -- } \label{fig:TQC-LT} 
Absolute value of correlation function at Fibonacci times for varying $\delta \theta = \pi - \theta$ and $L$. Here $\lambda = 1$ and random variables drawn from uniform distributions are $J \in [2, 8] $, $h^{z}_{i} \in [0, 0.5] $, $h^{x}_{i} \in [0, 0.03]$. We average over 500 disorder realizations and average over all states in the full Hilbert space for each disorder.}
\vspace{-0.2in}
\end{figure}

\newpage

\end{document}